# Wire twisting stiffness modelling with application in wire race ball bearings. Derivation of analytical formula and finite element validation

J. Aguirrebeitia, I. Martín, I. Heras, M. Abasolo, I. Coria

**Abstract**

Since Erich Franke produced the first wire race bearings in 1934, they have not been used profusely until these last years in applications such as computerized tomography, X-ray machines, wheels with direct drive…where low weight and inertia constraints are important. Accounting for the structural behaviour of the bearing, there exist a key phenomenon not present in other kind of more known bearings, which is the wire twisting under load; this wire twisting steers the force transmission among the bearing rings and the rolling elements. In this sense, for design and selection purposes, if a complete structural model of the bearing is to be done in an efficient way to assess bearing stiffness and load distribution over the rolling elements, the twisting stiffness of the wire has to be modelled properly. This work develops a simple analytical expression of that stiffness to be used in structural models, derived from an evidence-based deformation assumption at differential level for the section of the wire.

## 1. Introduction

Wire race ball bearings are elements designed to support rotatory components of machinery with weight and inertia constraints. In fact, and compared with conventional bearings, wire race bearings have a lightweight body, made of light materials as aluminium, and only the raceways, in the form of a “wire”or thin ring, are made of hardened steel. This fact makes these bearings lighter than conventional ones (up to 60% weight savings in manufacturers’ words). Fig. 1 shows the outline of a four contact point wire race ball bearing.

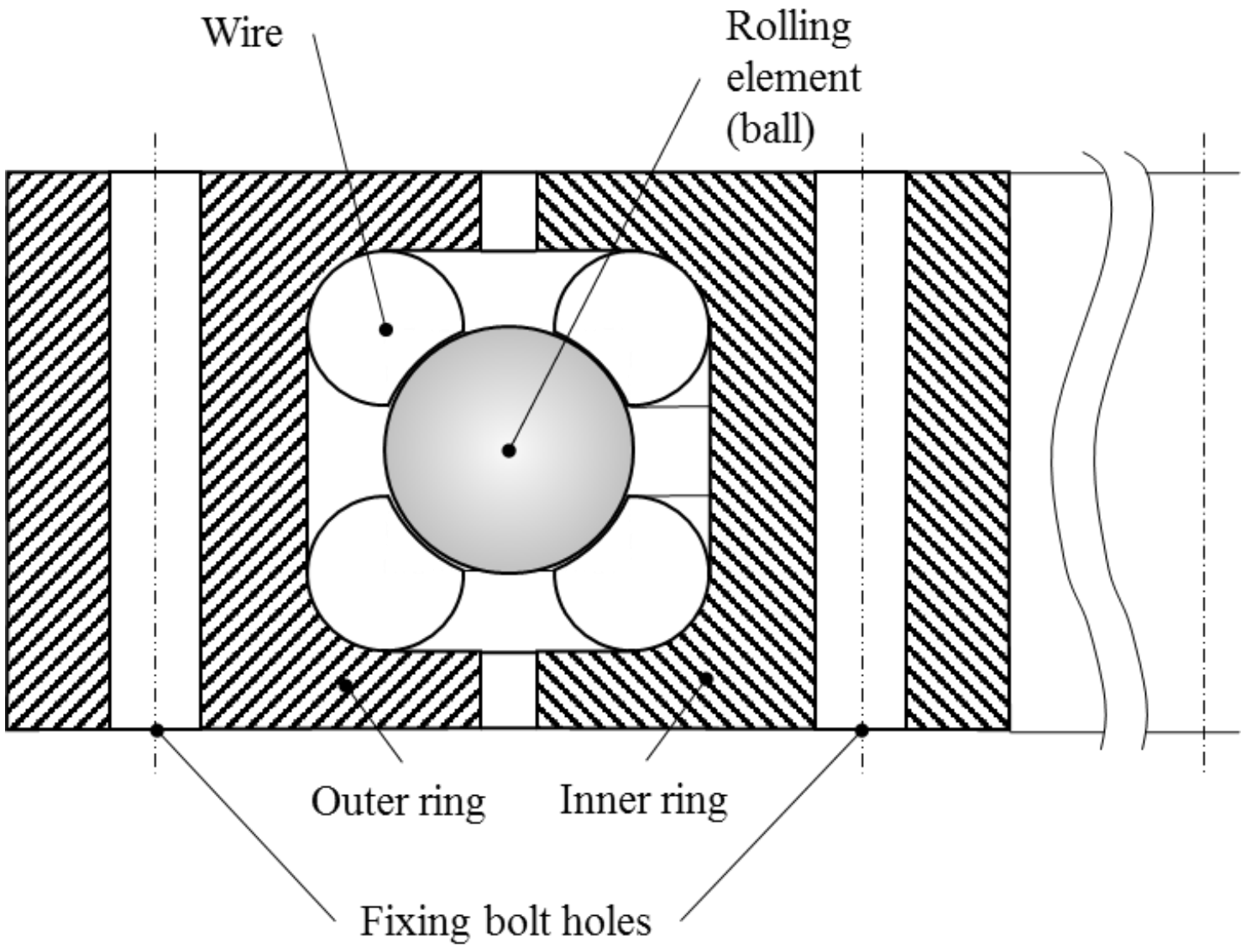

**Figure 1.** Wire race ball bearing

Little journal publications have been released about mechanical behaviour of this kind of bearings. Shan et al. [1] pub- lished a paper in which the preload of a rare kind of nonconformal wire bearing was assessed via an analytical approach. Later, Gunia and Smolnicki [2] made an interesting research about the stress distribution on wires and rolling elements in wire race bearings, in which the influence of some geometrical parameters was investigated. Martín et al. [3] studied the structural response of wire bearings under axial loads, finding some relationships between operational parameters and comparing the behaviour with conventional bearings.

The present work aims to deepen into the mechanical behaviour of this kind of bearing, trying to assess a phenomenon not present in conventional bearings: the behaviour of the wire raceway itself under load. In fact, the wires tend to go along with the rolling element due to the conformal "bite"machined in them to accommodate it; for such purpose, the rolling element needs to overcome the twisting moment of the wire and, therefore, the wire twists with respect to its circumferential axis [3,4] . Fig. 2 shows this effect.

The authors believe that the correct modelling of the twisting stiffness of the wire can be decisive to correctly assess the load distribution among the rolling elements and the raceways, and therefore the overall mechanical behaviour of the bearing. In addition, this twisting stiffness can be used to feed simplified models to assess the global mechanical behaviour of the bearings and their surrounding structures [5–7] .

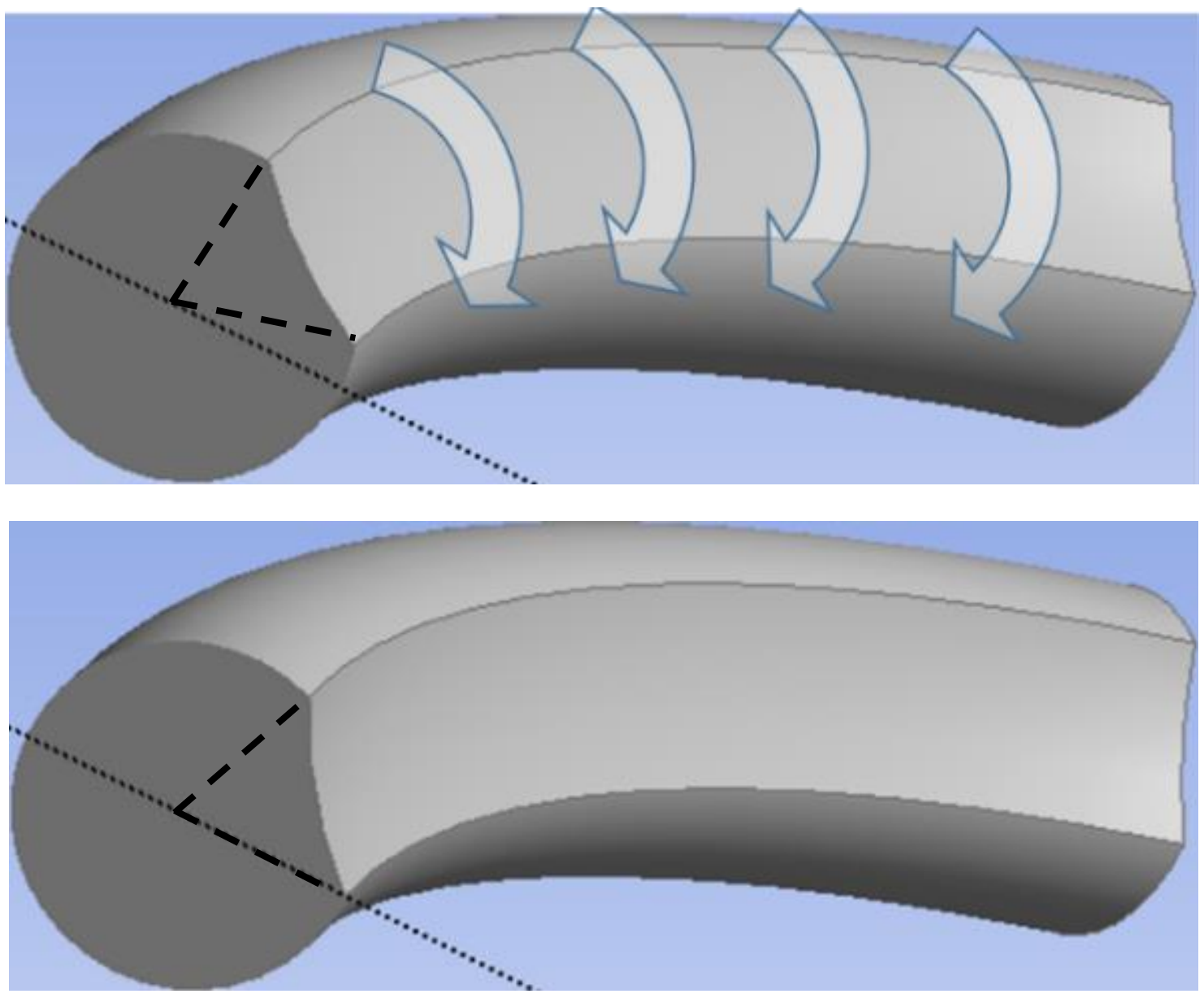

**Figure 2.** Wire twisting effect

In records as [8] , many formulae appear for elastic behaviour of many elements under several actions, and particularly the case of slender circular rings. Nevertheless, an analytical tool is to be derived here due to two reasons:

- Frequently, formulae in [9] are tricky and careful implementation is needed to avoid mistakes; apart from that, several errors have been reported in many forums along the years. This reflection does not mean that the authors misprize that valuable work; it is only about making sure that the basis of the oncoming work is well established.
- The observation of finite element analyses [3] , give some insight on how the ring behaves under the real boundary conditions and realistic elastic differential deformation assumptions can be done to accurately approach the real behaviour of the ring.

In this sense, as pointed out before, the wire twists with respect to its circumferential axis. In addition, concentrated loads due to ball-raceway contacts exert a quasi-distributed twisting moment, in such a way that circumferential “fibres” do not bend noticeably and they only deform circumferentially.

**2. Twisting stiffness of a circular wire with circular section**

Figure 3 shows a wire sector with a span angle $\beta = 2\pi/Z$ corresponding to the span of one rolling element, being Z the rolling element number. This sector has a cyclic-symmetry condition under axial load [3], in such a way that the rest of the sectors behave the same. In this point, let us assume first that the section is circular, and later the real shape of the ring will be considered.

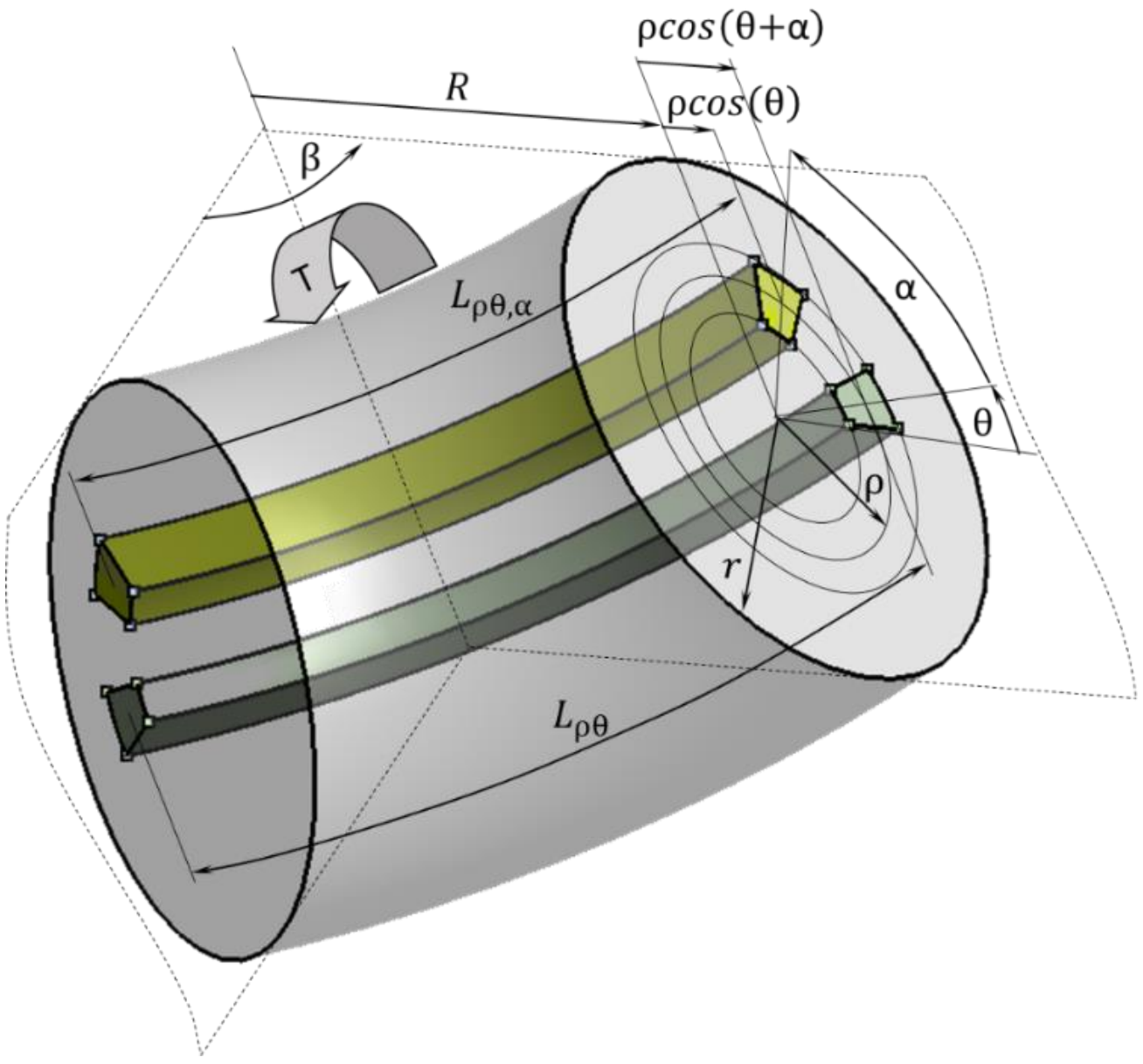

**Figure 3.** Wire sector corresponding to one rolling element

It will be assumed that a differential element located in polar coordinates $\{\rho, \theta\}$ with length$L_{\rho\theta}$, will vary its length to $L_{\rho\theta,\alpha}$ when the wire twists an angle $\alpha$ due to a twisting moment:

$$\begin{aligned} L_{\rho\theta} &= \beta\big(R + \rho cos(\theta)\big) \\ L_{\rho\theta,\alpha} &= \beta\big(R + \rho cos(\theta + \alpha)\big) \\ \Delta L_{\rho\theta,\alpha} = L_{\rho\theta,\alpha} - L_{\rho\theta} &= \beta\rho\big(cos(\theta + \alpha) - cos(\theta)\big) \end{aligned} \tag{1}$$

The differential force needed to perform this length variation can be expressed as the tractive-compressive stiffness differential constant $dK$ multiplied by the length variation $\Delta L_{\rho\theta,\alpha}$ as follows:

$$\begin{aligned} dF_{\rho\theta,\alpha} &= dK_{\rho\theta}\Delta L_{\rho\theta,\alpha} \\ dK_{\rho\theta} &= \frac{EdA_{\rho\theta}}{L_{\rho\theta}} \\ dA_{\rho\theta} &= \rho d\rho d\theta \end{aligned} \tag{2}$$

And the virtual work contribution of that differential force along the length variation:

$$\begin{aligned} dW_{\rho\theta,\alpha} &= dF_{\rho\theta,\alpha}\Delta L_{\rho\theta,\alpha} \\ dW_{\rho\theta,\alpha} &= E\frac{\left(\Delta L_{\rho\theta,\alpha}\right)^2}{L_{\rho\theta}}\rho d\rho d\theta \\ dW_{\rho\theta,\alpha} &= \beta E\frac{\big(cos(\theta + \alpha) - cos(\theta)\big)^2}{\big(R + \rho cos(\theta)\big)}\rho^3 d\rho d\theta \end{aligned} \tag{3}$$

The sum of the differential virtual work for all the differential elements over the section of the wire must be equal to the virtual work done by the twisting moment $T$ along the angle $\alpha$. From that equality, the twisting moment can be expressed:

$$T = \frac{\beta E}{\alpha} \iint_A \frac{\left(cos(\theta + \alpha) - cos(\theta)\right)^2}{\left(R + \rho cos(\theta)\right)} \rho^3 d\rho d\theta \tag{4}$$

For small twisting angles $\alpha$ and for wire radius *R* much larger than section radius *r*, the previous expression can be simplified:

$$\begin{gathered} \left(cos(\theta + \alpha) - cos(\theta)\right)^2 \cong \alpha^2 sin^2(\theta) \\ R + \rho cos(\theta) \cong R \\ T \cong \frac{\alpha \beta E}{R} \iint_A \rho^3 sin^2(\theta) d\rho d\theta \end{gathered} \tag{5}$$

This last equation allows to define the twisting stiffness constant as the twisting moment *T* over the twisting angle $\alpha$:

$$K_T = \frac{\beta E}{R} \iint_A \rho^3 sin^2(\theta) d\rho d\theta \tag{6}$$

For a circular section, the integral in (6) is easily solved since the integration limits are constant and independent:

$$\begin{gathered} K_T = \frac{\beta E}{R} \int_0^{2\pi} sin^2(\theta) \left[ \int_0^r \rho^3 d\rho \right] d\theta \\ K_T = \beta E \frac{\pi r^4}{4R} \end{gathered} \tag{7}$$

And considering the definition for the span angle $\beta$ as a function of the ball number in the bearing:

$$K_T = \frac{E r^4}{Z R} \left( \frac{\pi^2}{2} \right) \tag{8}$$

Which is a simple expression easily applicable. For typical values of a wire bearing [3, 4]:

$$\begin{gathered} R = 227\, mm \\ r = 3{,}3\, mm \\ Z = 82\, balls \\ E = 210{,}000\, MPa \\ K_T = 6602\, \frac{N \cdot mm}{rad} \end{gathered} \tag{9}$$

### 3. Twisting stiffness of a circular ring with real wire race section

The wire for bearing applications has a circular section from which another non-centred circle is subtracted to finally give a section like in Fig. 2. Fig. 4 shows the geometric parameters of the section expressed in polar coordinates.

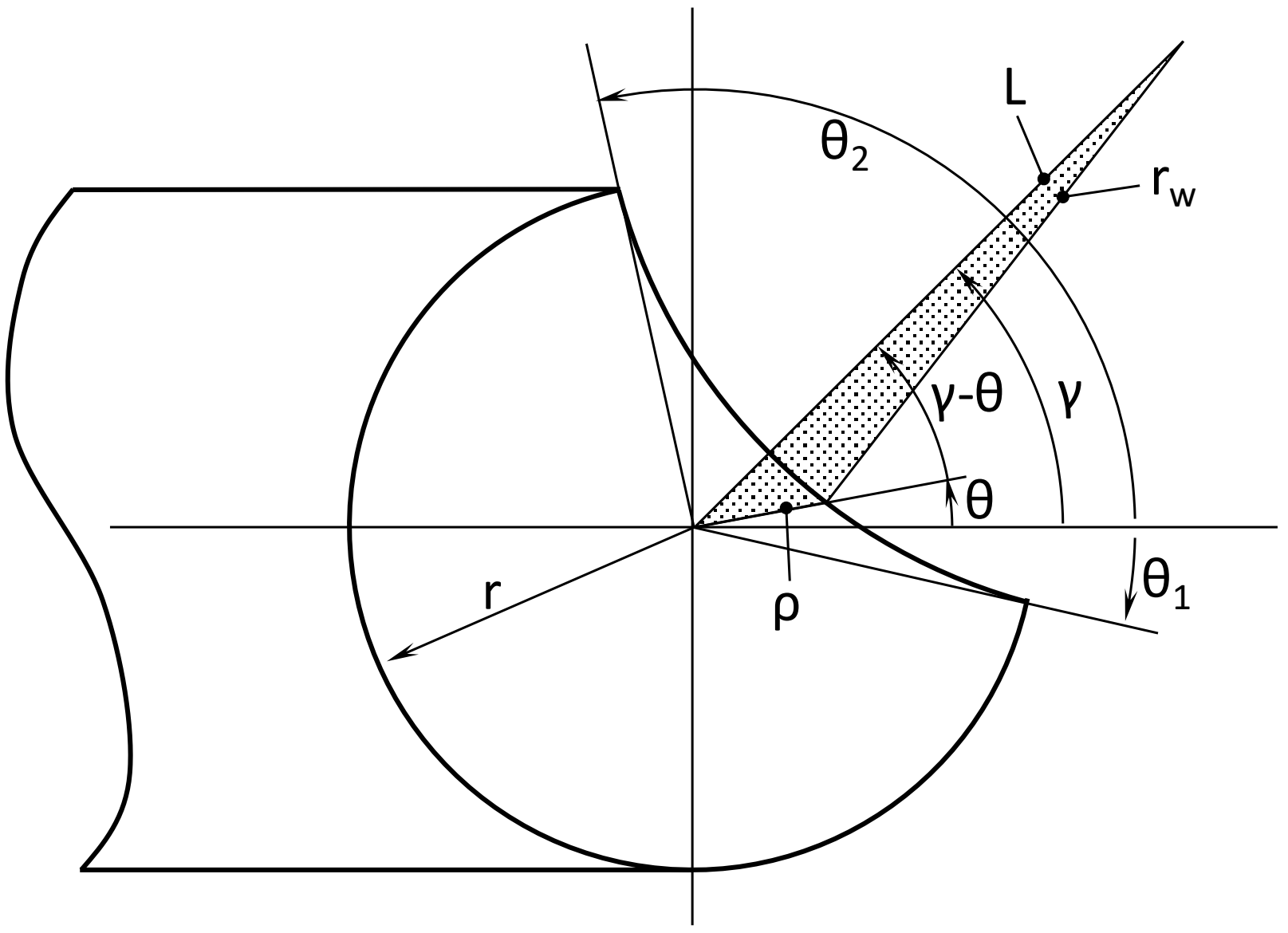

**Figure 4.** Geometric parameters for real wire section

The integral in (6) must be split into two parts. In fact, definite Integral $I_1$ in (10) can be easily solved as in (7) since the integration limits are independent for the two polar variables, but integral $I_2$ must be solved for $\rho(\vartheta)$ from $\vartheta_1$ to $\vartheta_2$.

$$
\begin{aligned}
I &= \iint_A \rho^3 sin^2(\theta) d\rho d\theta = I_1 + I_2 \\
I_1 &= \int_{\theta_2}^{2\pi+\theta_1} sin^2(\theta) \left[\int_0^r \rho^3 d\rho\right] d\theta \\
I_2 &= \int_{\theta_1}^{\theta_2} sin^2(\theta) \left[\int_0^{\rho(\theta)} \rho^3 d\rho\right] d\theta
\end{aligned} \tag{10}
$$

Then, integral $I_1$:

$$
\begin{aligned}
I_1 &= \int_{\theta_2}^{2\pi+\theta_1} sin^2(\theta) \left[\int_0^r \rho^3 d\rho\right] d\theta = \frac{r^4}{4}\left[\frac{\theta}{2} - \frac{sin(2\theta)}{4}\right]_{\theta_2}^{2\pi+\theta_1} \\
I_1 &= \frac{r^4}{4}\left(\pi + \frac{(\theta_1 - \theta_2)}{2} + \frac{\big(sin(2\theta_2) - sin(2\theta_1)\big)}{4}\right)
\end{aligned} \tag{11}
$$

For integral $I_2$ first the equation of a non-centered circumference is derived in the form of $\rho(\vartheta)$. Applying the law of the cosine to the shaded triangle in Figure 4:

$$
r_w^2 = L^2 + \rho(\theta)^2 - 2L\rho(\theta)cos(\gamma - \theta) \tag{12}
$$

Solving for the first value of $\rho(\vartheta)$ which fulfills (12),

$$
\rho(\theta) = Lcos(\gamma - \theta) - \sqrt{r_w^2 - L^2 sin^2(\gamma - \theta)} \tag{13}
$$

Therefore, integral $I_2$ can be solved as:

$$I_2 = \frac{1}{4}\int_{\theta_1}^{\theta_2} sin^2(\theta)\rho(\theta)^4 d\theta$$

$$I_2 = \frac{r^4}{4}\int_{\theta_1}^{\theta_2} sin^2(\theta)\left(\frac{L}{r}cos(\gamma-\theta) - \sqrt{\left(\frac{r_w}{r}\right)^2 - \left(\frac{L}{r}\right)^2 sin^2(\gamma-\theta)}\right)^4 d\theta \qquad (14)$$

Which must be solved numerically for $r_w/r$, $L/r$ and $\gamma$. Regarding the integration limits for both integrals, depending on geometric parameters on Figure 4, values for $\vartheta_1$ and $\vartheta_2$ can be derived, doing $\rho(\vartheta)=r$ in (15):

$$\theta_1 = \gamma - arccos\left(\frac{1+\left(\frac{L}{r}\right)^2 - \left(\frac{r_w}{r}\right)^2}{2\left(\frac{L}{r}\right)}\right)$$
$$\theta_2 = \gamma + arccos\left(\frac{1+\left(\frac{L}{r}\right)^2 - \left(\frac{r_w}{r}\right)^2}{2\left(\frac{L}{r}\right)}\right) \qquad (15)$$

In this point a DoE is planned to obtain an engineering formula in order to make a reasonable approximation for integrals in (10). The DoE is fully factorial and it has been done for the values in Table 1, within the limits adopted generally for the geometric parameters. Applicable values for $\gamma$ in Table 1 are *π/4*, *3π/4*, *5π/4* and *7π/4*.

From Table 1 it can be reasonably concluded that the value of the Integral has a strong linear relationship with parameter ($L/r-r_w/r$) and second order dependences can be mispriced. The following equation can be derived via Least Squares, forcing the integral to be *π/4* for $L/r-r_w/r=1$:

$$I \approx r^4\left(\frac{\pi}{4} - 0{,}36\left[1-\left(\frac{L}{r}-\frac{r_w}{r}\right)\right]\right) \qquad (16)$$

Table 1. Integral (10) for geometrical parameters within the common design range, valid for $\gamma=\pi/4$, $\gamma=3\pi/4$, $\gamma=5\pi/4$ and $\gamma=7\pi/4$:

| $r_w/r$ | $L/r$ | $L/r- r_w/r$ | $I/r^4$ |
|---|---|---|---|
| 2 | 2,25 | 0,25 | 0,522088805 |
| 2,5 | 2,75 | 0,25 | 0,513905797 |
| 3 | 3,25 | 0,25 | 0,503645074 |
| 2 | 2,5 | 0,5 | 0,609857837 |
| 2,5 | 3 | 0,5 | 0,604207922 |
| 3 | 3,5 | 0,5 | 0,600254386 |
| 2 | 2,75 | 0,75 | 0,708211997 |
| 2,5 | 3,25 | 0,75 | 0,705624706 |
| 3 | 3,75 | 0,75 | 0,703774932 |
| 2 | 3 | ≥1 | π/4 |
| 2,5 | 3,5 | ≥1 | π/4 |
| 3 | 4 | ≥1 | π/4 |

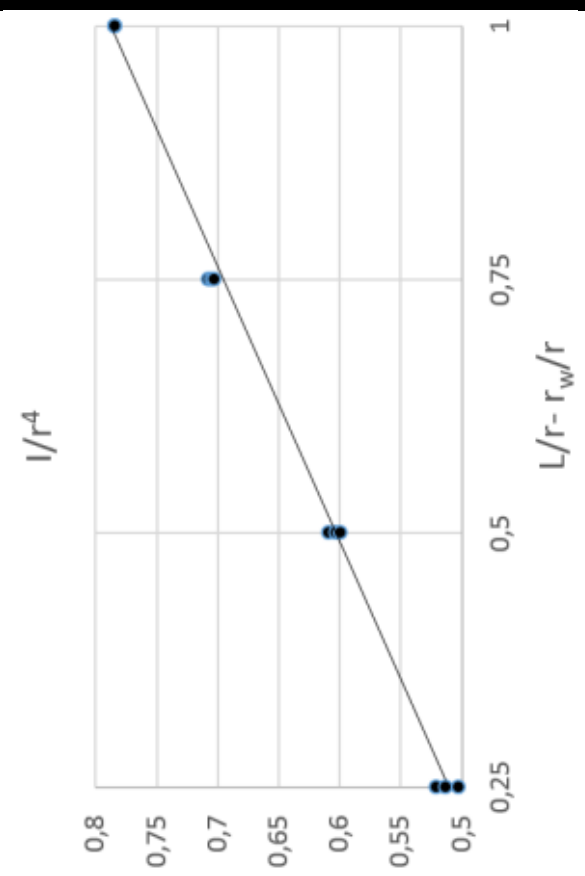

Then, the stiffness constant can be expressed as:

$$K_T = \beta E \frac{r^4}{R}\left(\frac{\pi}{4} - 0{,}36\left[1 - \left(\frac{L}{r} - \frac{r_w}{r}\right)\right]\right)$$
$$K_T = \frac{Er^4}{ZR}\left(\frac{\pi^2}{2} - 0{,}72\pi\left[1 - \left(\frac{L}{r} - \frac{r_w}{r}\right)\right]\right) \quad (17)$$

**4. Finite element modelling of wires and analytical-numerical correlation**

*4.1. Simplified load cases*

As a first step in order to check the validity of the formulae, the exactly same cases developed before are reproduced with finite elements using ANSYS®, with the same way of imposing the torque. Table 2 shows the models and systems of Loads & Boundary conditions used for the case of circular section and the same for a real wire section. Stiffness constants have been derived for a twisting angle interval [α=0…0,1 rad] and also for the immediacies of α=0. For the noncircular case, stiffness coefficients have been computed for positive and negative twisting angles. The formulation developed above behaves really good for the circular case, and it behaves reasonably good for the noncircular case as Table 2 shows. For the latter case, the behavior of γ=45º has been shown only, since the case γ=135º behaves the same, but for opposite sign of the twisting angle.

*4.2. A more realistic load case*

In order to deepen into the validity of the stiffness constant calculated by the analytical approach, a more realistic load case is considered now. In this case, a rigid support for the wire has been considered, as well as a rigid ball as loading element. The ball approaches the wire in such a direction that finally the contact materialize almost in the edge of the wire, which is the situation for which the distribution of applied torque is the furthest possible from the theoretical assumption. Analysis have been done for friction coefficients μ=0.0 and μ=0.1 for the three contact pairs defined in the model, being the latter case a very usual one in slewing bearings [9,10]. Table 3 shows the model and the results.

**5. Results and Discussion**

Tables 2 and 3 show the fitness of the formulae developed in this work. Results for simplified load cases in Table 2 show that in a circular section, accuracy of Eq. (8) when comparing with Finite Element Models is near 100%. For the case study of a real wire section, the accuracy of Eq. (17) is around 98%. Results for the realistic load case in Table 3 show that Eq. (17) fits adequately the stiffness curve both for the frictionless and frictional cases. For both cases, the resulting moment of the applied forces in the wire were computed in the Finite Element code, and compared with Eq. (17) .

Regarding the results given in the tables, some aspects must be further commented. In Table 2, K Origin refers to the stiffness of the first load step of the FE model, i.e. considering the initial undeformed geometry. As the analytical model is based on this undeformed geometry, K Origin coincides with the analytical stiffness, and its value is the same for any twisting direction. However, as the twisting angle increases, the wire section rotates and therefore its stiffness varies. In contrast to the analytical model, the FE model accounts for this phenomenon (large displacements), and for that reason FE and analytical stiffness values are slightly different. Moreover, the FE results are different for both twisting directions ($T^+$ and $T^-$). Obviously, this effect only takes place for the real wire section case. Table 2 also shows that both the axisymmetric and cyclic symmetric models provide the same results, even though the load was applied in different ways as explained in Section 4.1. Despite all these considerations, the discrepancies between the analytical and FE model are negligible, as outlined in the previous paragraph.

In reference to Table 3, the contact pressure plot on the right shows that for the final displacement condition, the load acts on the lower edge of the wire, as mentioned in Section 4.2 (the left plot corresponds the first load step). As the displacement imposed to the ball is the same for $\mu$= 0.0 and $\mu$= 0.1 cases, so is the displacement (and twisting angle) of the wire. As the stiffness of the wire is the same regardless of the contact behaviour, the twisting moments are also the same for the frictionless and frictional cases. However, in the $\mu$= 0.1 case friction forces appear in the wire-support contacts, so the ball-wire contact force is larger than in the $\mu$= 0.0 case in order to fulfil the static equilibrium. Anyway, Table 3 shows that the analytical stiffness approaches satisfactorily FE results for both frictionless and frictional contact conditions.

Table 2. Results comparison between models with ideally distributed moments.

| ***Case*** | ***Model*** | ***Loads & Boundary Conditions*** | ***Stiffness Constant [N.mm/rad]*** |
|---|---|---|---|
| Section: Circular<br>R: 227 mm<br>r: 3,3 mm<br>Z: 82 balls<br>E: 210 GPa | 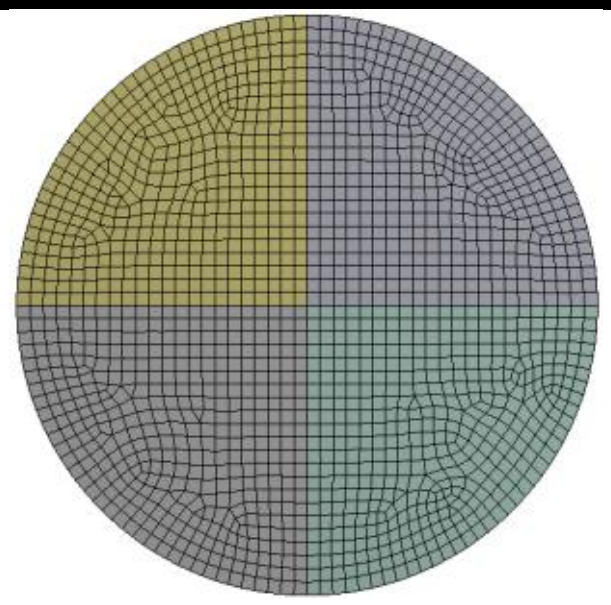<br>FE Axisymmetric | 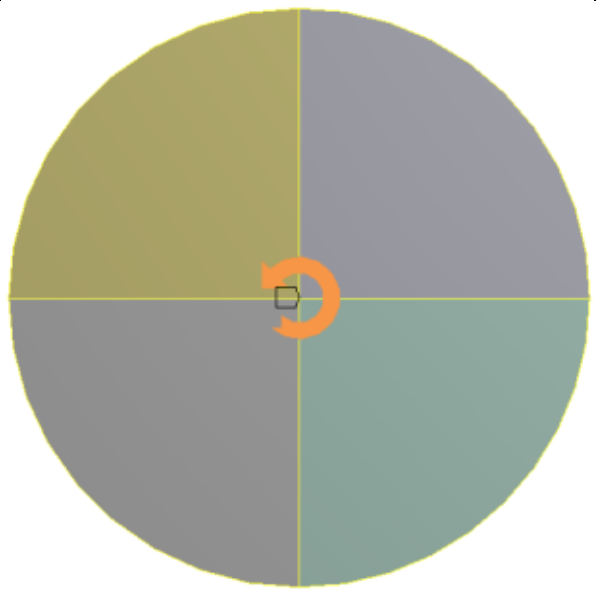<br>Imposed twist to section | 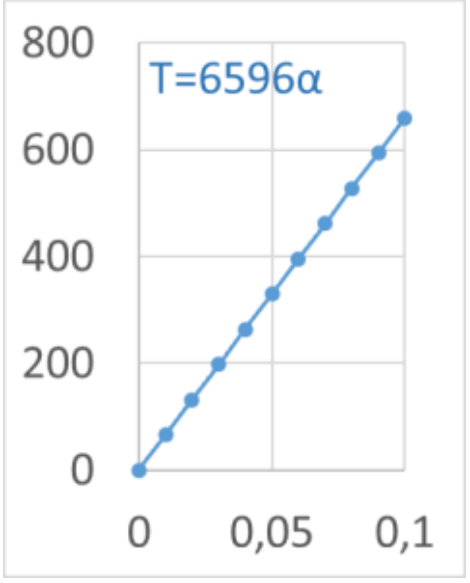<br>$K^{Origin}$=6602 |
| | 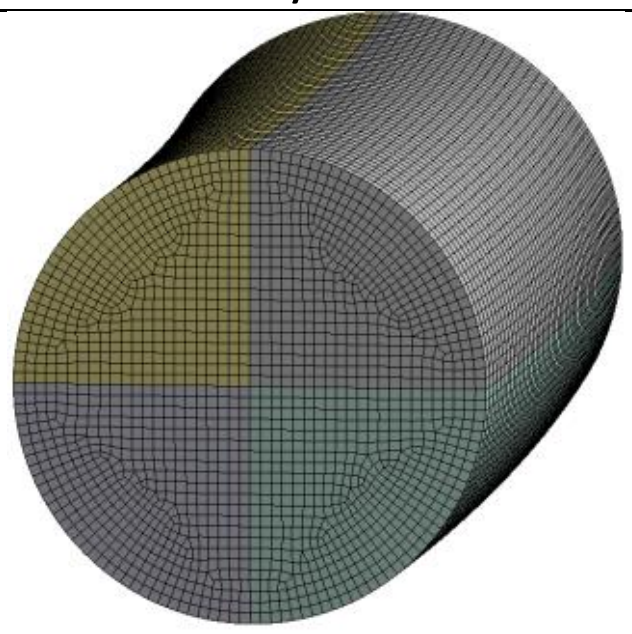<br>FE Cyclic Symmetry | 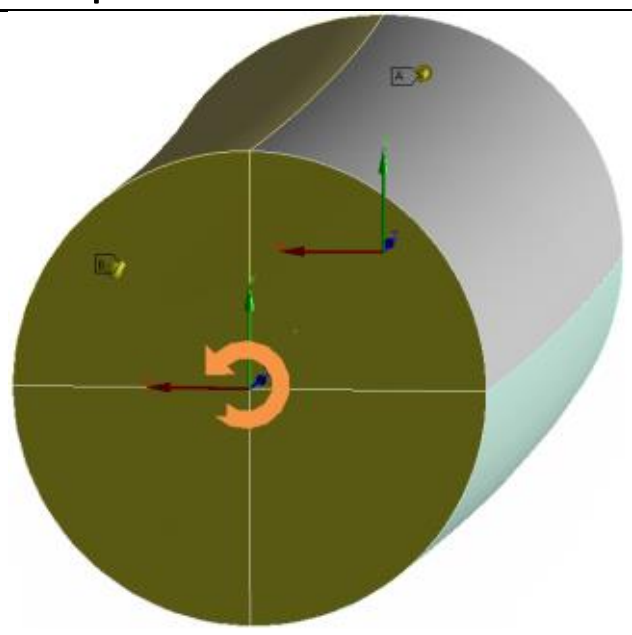<br>Imposed twist on sides | 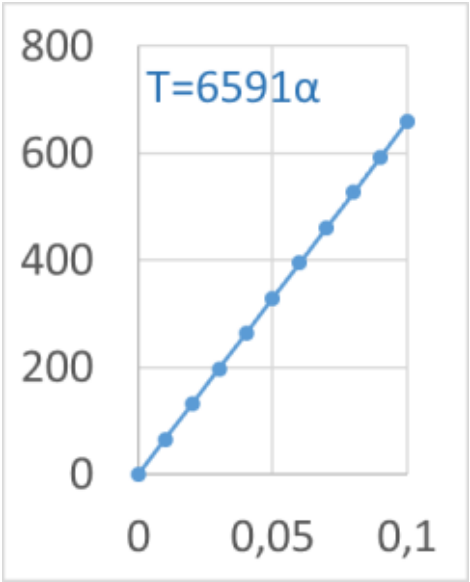<br>$K^{Origin}$=6598 |
| | Analytical (8) | Analytical (8) | K=6602 |
| Section: Real<br>R: 227 mm<br>r: 3,3 mm<br>$r_w$/r: 3,0<br>L/r: 3,5<br>γ: 45º<br>Z: 82 balls<br>E: 210 GPa | 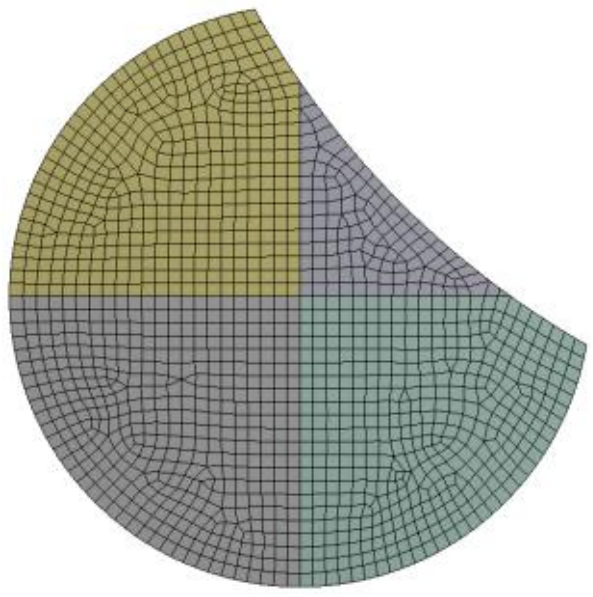<br>FE Axisymmetric | 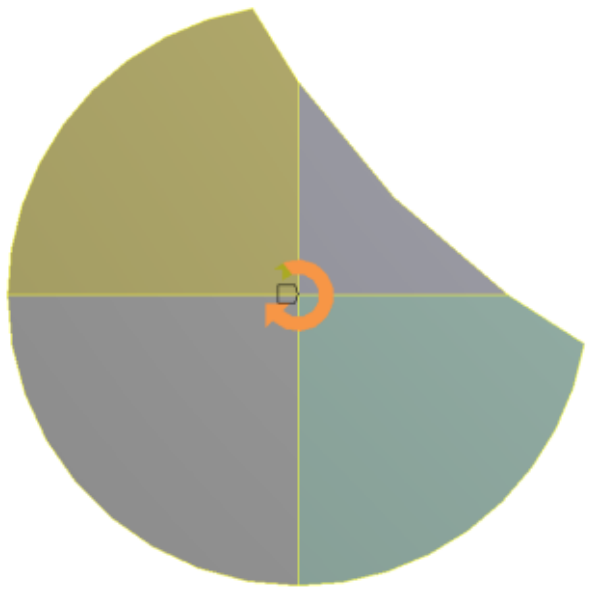<br>Imposed twist to section | 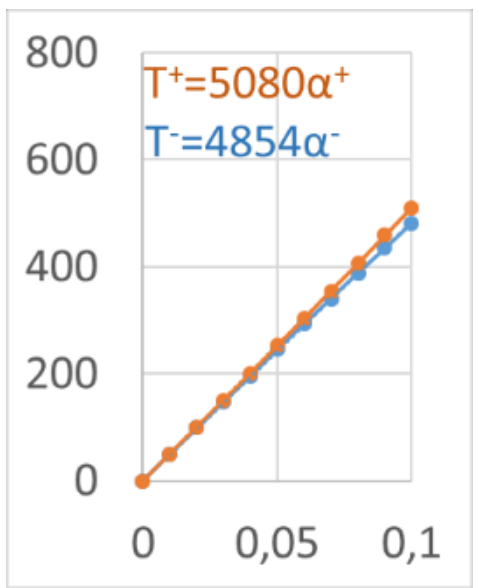<br>$K^{Origin}$=4975 |
| | 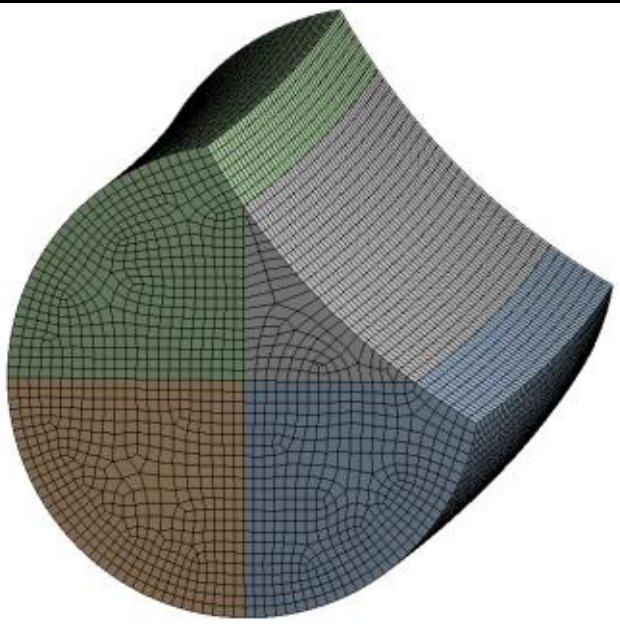<br>FE Cyclic Symmetry | 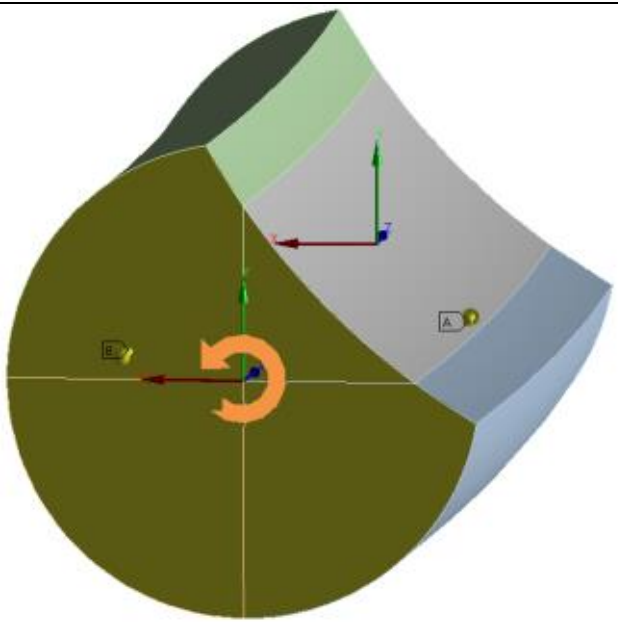<br>Imposed twist on sides | 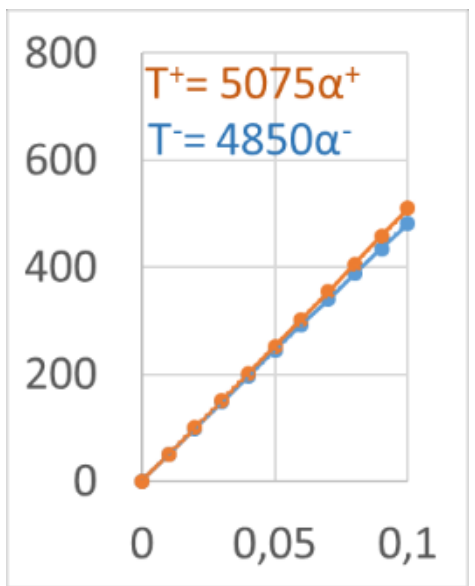<br>$K^{Origin}$=4970 |
| | Analytical (12,15) | Analytical (12,15) | K=5046 |
| | Analytical (18) | Analytical (18) | K=5089 |

Table 3. Results comparison between models with realistically distributed moments.

| ***Cyclic Symmetric FE Model*** | ***Loads & Boundary Conditions*** |
|---|---|
| 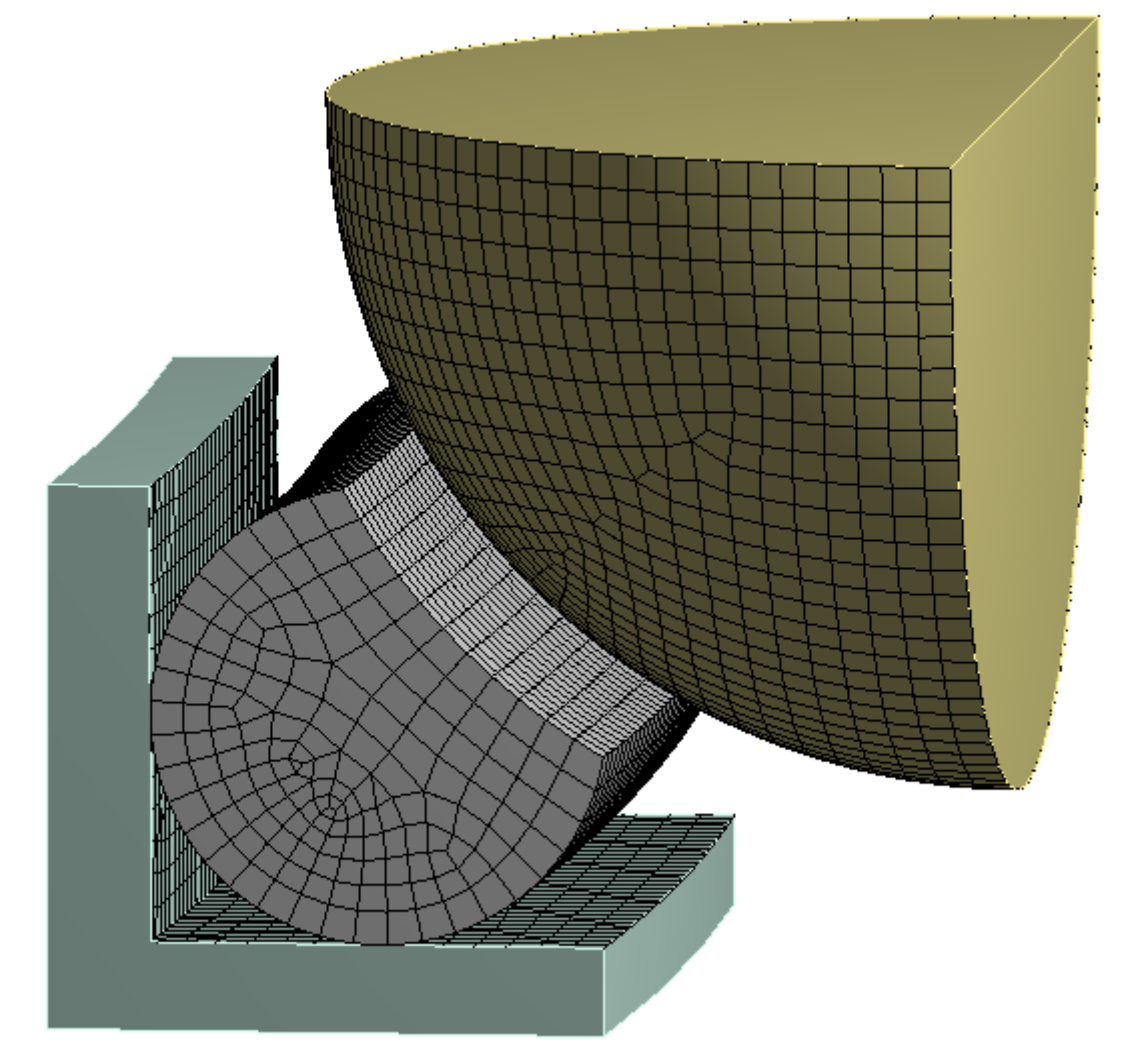 | 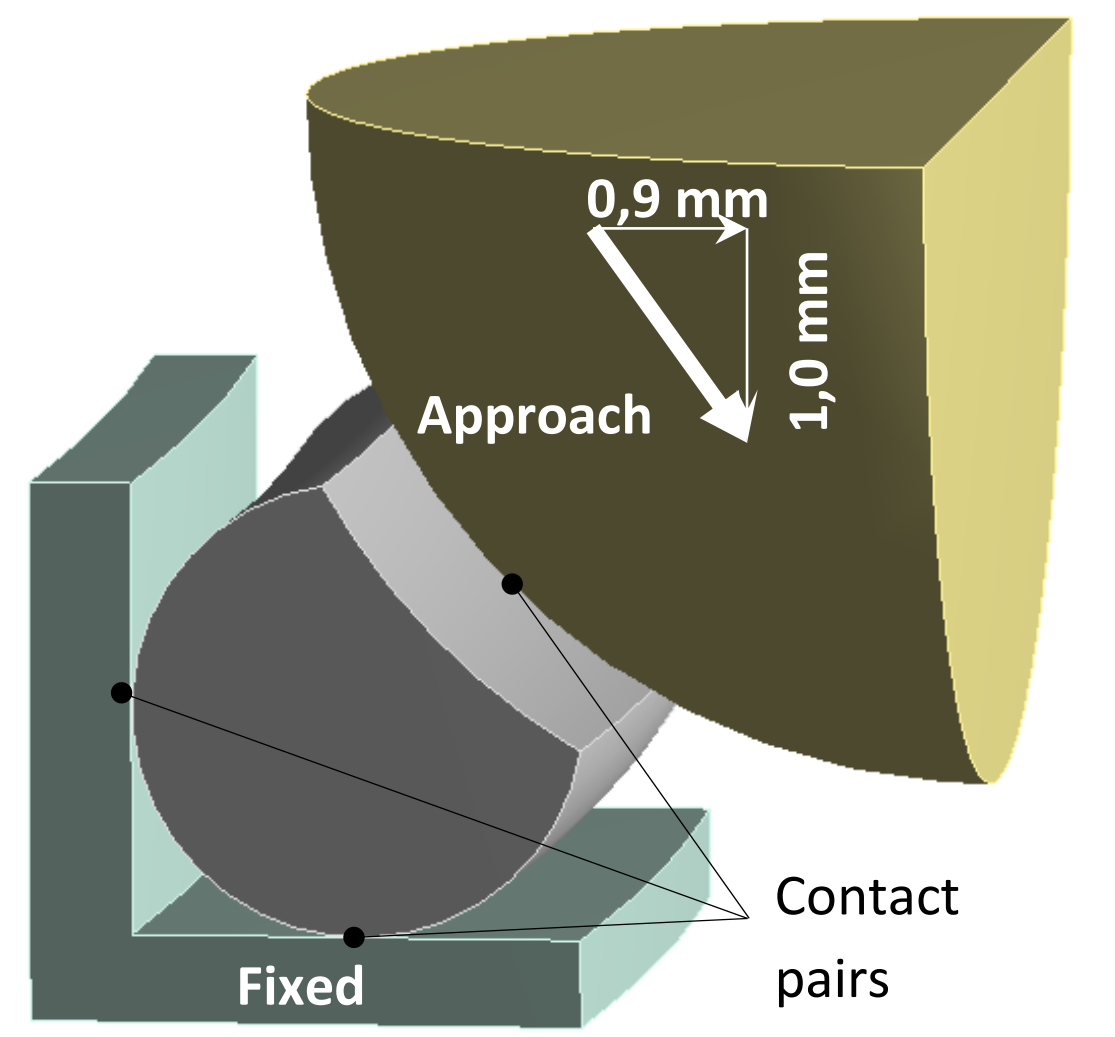 |

***Case μ=0.0***

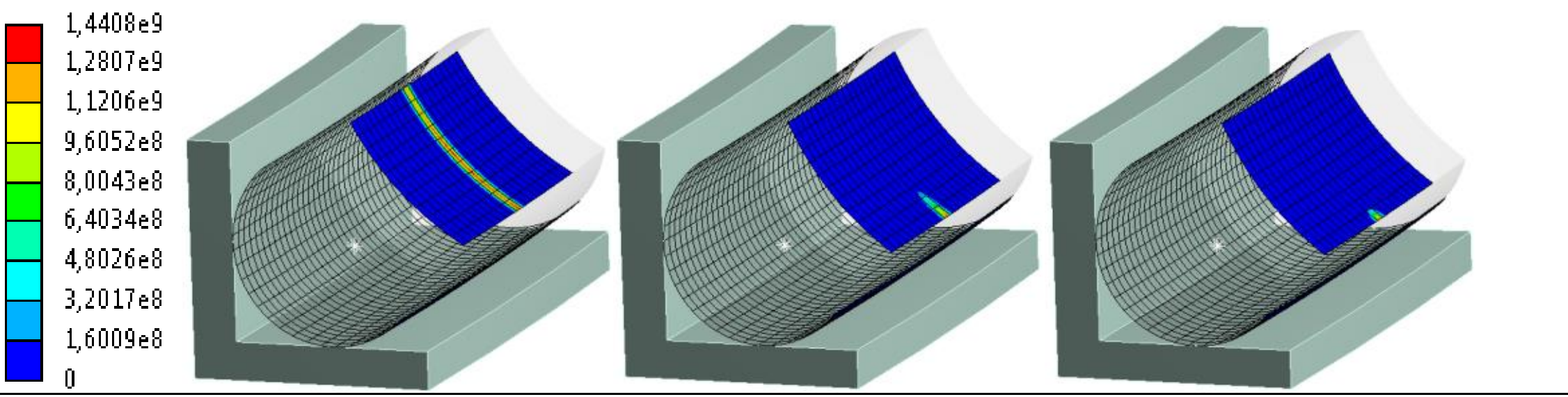

Pressure distribution during loading

***Case μ=0.0. Last load step***

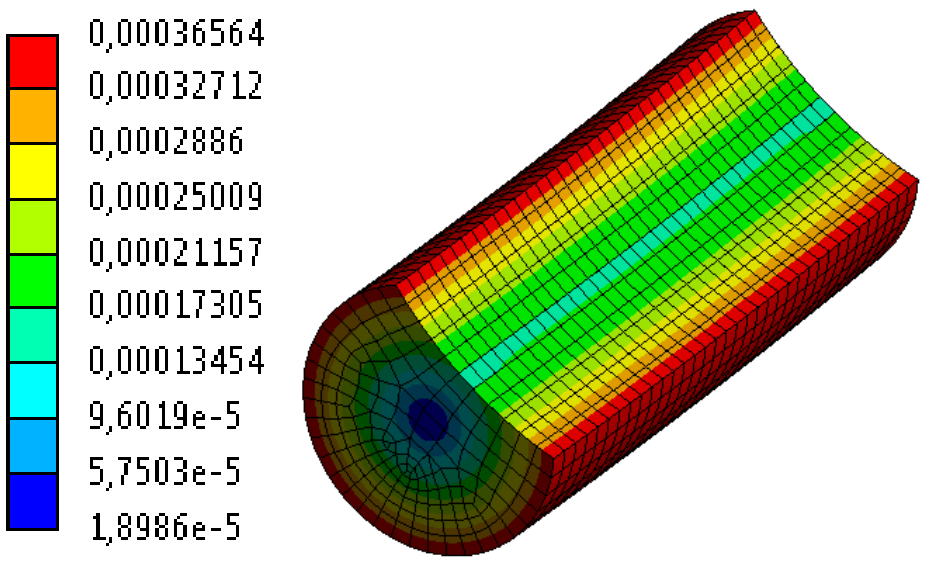

Displacement Distribution

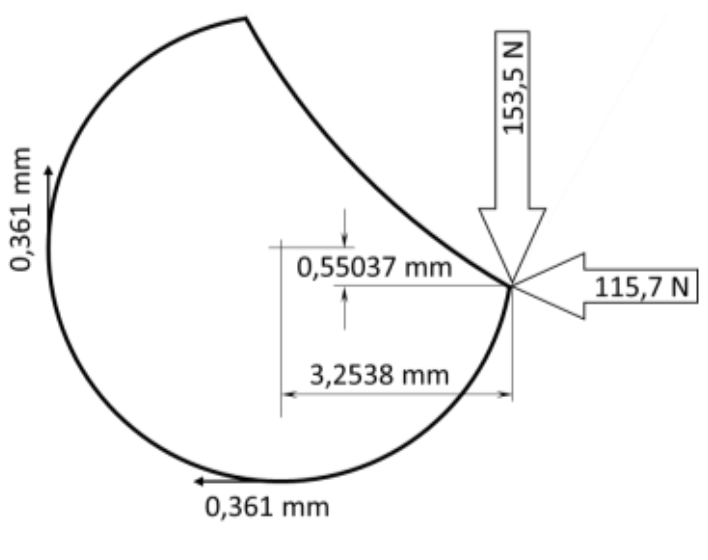

Load vs. Displacement

Moment

T= 153,5·3,2538+ 115,7·0,55037= 564 N.mm

Angle

α=0,361/3,3=0,11 rad

Stiffness constant

**K=5146 N.mm/rad**

Stiffness Constant

***Cases μ=0.0 and μ=0.1. Moment vs. angle for 10 load steps***

| Angle [rad] | Moment FE-μ=0.0 [N·mm] | Moment Analytic (17) [N·mm] |
|---|---|---|
| 0,0000 | 0 | 0 |
| 0,0117 | 66 | 60 |
| 0,0234 | 130 | 119 |
| 0,0349 | 191 | 178 |
| 0,0465 | 253 | 237 |
| 0,0582 | 313 | 296 |
| 0,0698 | 369 | 355 |
| 0,0815 | 424 | 415 |
| 0,0923 | 474 | 470 |
| 0,1010 | 520 | 514 |
| 0,1095 | 564 | 557 |

| Angle [rad] | Moment FE-μ=0.1 [N·mm] | Moment Analytic (17) [N·mm] |
|---|---|---|
| 0,0000 | 0 | 0 |
| 0,0058 | 32 | 29 |
| 0,0115 | 66 | 58 |
| 0,0172 | 99 | 87 |
| 0,0230 | 131 | 117 |
| 0,0286 | 163 | 145 |
| 0,0343 | 194 | 174 |
| 0,0400 | 225 | 204 |
| 0,0457 | 255 | 233 |
| 0,0515 | 284 | 262 |
| 0,0574 | 315 | 292 |

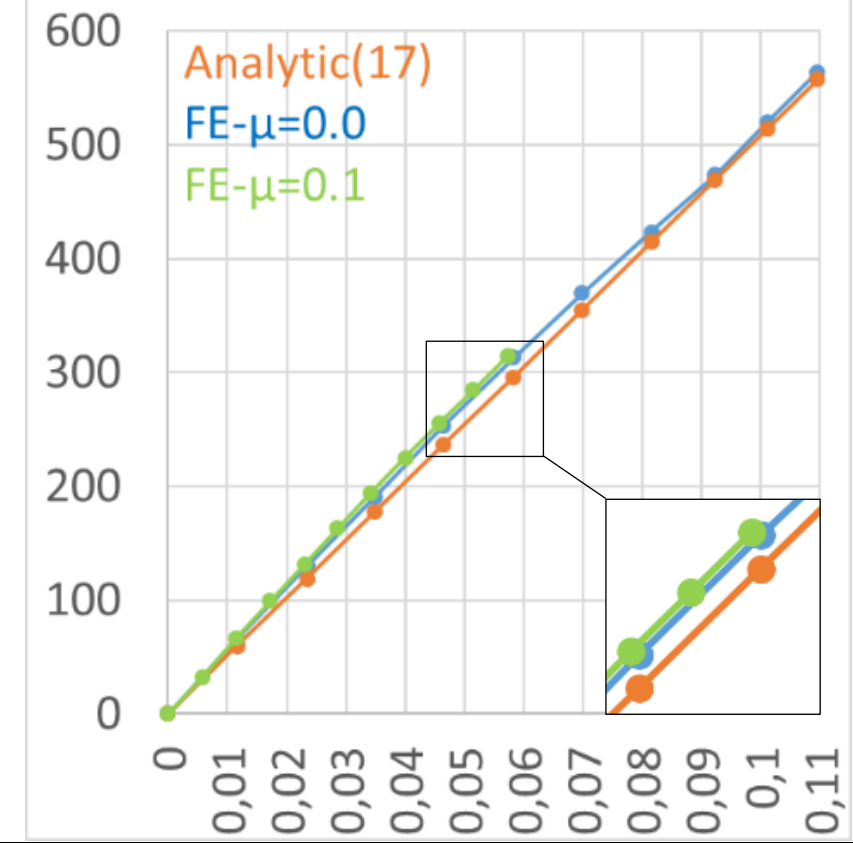

## Concluding remarks

In this work, an engineering formula was derived to model the twisting stiffness of wires present in wire race bearings. For that, an evidence-based deformation assumption was done at differential level for the section of the wire, and after its integration and further mapping, an easy-to-use formula was tuned as a function of the principal geometrical parameters of the wire. The results attest that wire twisting stiffness can be modelled quite accurately with that formula, and eventually this stiffness can be included in a simplified model of the whole bearing to account for bearing stiffness and load distribution.

## Acknowledgements

The authors want to acknowledge the financial support of the Spanish Ministry of Economy and Competitiveness [grant number DPI2017-85487-R (AEI/FEDER , UE)], the Basque Government [grant number IT947-16 ] and finally, the Basque Bearing Manufacturer Iraundi Bearings S.A. for the help in the geometrical definition of the wires and many background items of the research.